\documentclass[aps,preprint,pre]{revtex4}

\usepackage{psfrag}

\usepackage{epsfig}
\usepackage{xspace}
\usepackage{tabularx}

\usepackage{amssymb}
\usepackage{amsmath}
\usepackage{amsbsy}

\def\bra{\langle}
\def\ket{\rangle}
\newcommand{\ch}{\chi\rule[-1ex]{0mm}{2ex}}
\DeclareMathSymbol{\MMM}{\mathbin}{AMSb}{'115}
\DeclareMathSymbol{\PPP}{\mathbin}{AMSb}{'120}
\DeclareMathSymbol{\DDD}{\mathbin}{AMSb}{'104}
\DeclareMathSymbol{\GGG}{\mathbin}{AMSb}{'107}
\DeclareMathSymbol{\SSS}{\mathbin}{AMSb}{'123}
\DeclareMathSymbol{\ZZZ}{\mathbin}{AMSb}{'132}

\def\FD{\hbox{\sffamily FD}\xspace}
\def\LB{\hbox{\sffamily LB}\xspace}
\def\EX{\hbox{\sffamily EX}\xspace}
\def\DM{\hbox{\sffamily DM}\xspace}
\def\SA{\hbox{\sffamily SA}\xspace}
\def\GF{\hbox{\sffamily GF}\xspace}
\def\corfct{\hbox{\sffamily S}_2}
\def\mum{\;\mu{\rm m}}
\def\mm{\;{\rm mm}}
\def\mD{\;{\rm mD}}

\begin{document}

\title{Lattice-Boltzmann and finite-difference simulations for the
  permeability for three-dimensional porous media}

\author{C.~Manwar${\rm t}^1$, U.~Aaltosalm${\rm i}^2$, A.~Kopone${\rm
    n}^2$, R.~Hilfe${\rm r}^{1,3}$, and J.~Timone${\rm n}^2$} 

\affiliation{$\ ^1$ ICA-1, Universit\"at
  Stuttgart, 70569 Stuttgart, Germany\\
  $\ ^2$ Department of Physics, University of Jyv\"askyl\"a, P.O. Box
  35, FIN-40351 Jyv\"askyl\"a, Finland\\
  $\ ^3$ Institut f\"ur Physik, Universit\"at Mainz, 55099 Mainz, Germany} 
\date{\today}

\begin{abstract}

Numerical micropermeametry is performed on three dimensional porous
samples having a linear size of approximately 3 mm and a resolution of 7.5
$\mu$m. One of the samples is a microtomographic image of Fontainebleau
sandstone. Two of the samples are  stochastic reconstructions with
the same porosity, specific surface area, and two-point correlation
function as the Fontainebleau sample. 
The fourth sample is a physical model which mimics the
processes of sedimentation, compaction and diagenesis of 
Fontainebleau sandstone.
The permeabilities of these samples are determined by numerically
solving at low Reynolds numbers the appropriate Stokes equations in the
pore spaces of the samples. The physical
diagenesis model appears to reproduce the permeability of the real
sandstone sample quite accurately, while the permeabilities of the
stochastic reconstructions deviate from the latter by at least an order
of magnitude.
This finding confirms earlier qualitative predictions based on
local porosity theory.
Two numerical algorithms were used in these
simulations. One is based on the lattice-Boltzmann method, and the other
on conventional finite-difference techniques. The accuracy of these two
methods is discussed and compared, also with experiment. 
\bigskip
\begin{tabbing}
  PACS: \= 81.05.Rm, 83.85.Pt, 61.43.Gt
\end{tabbing}
{\tt Phys.Rev.E (2002), in print}
\end{abstract}
\maketitle

\section{Introduction}
Almost all investigations of porous media focus on the prediction of
effective material properties such as fluid permeability, electric or
thermal conductivity, or elastic constants \cite{sah95,hil95d}. The
knowledge or at least a reliable prediction of these properties is of
great interest in a wide field of technical applications ranging from
petroleum engineering \cite{KBDHLPS99,lak89}, to paper manufacturing
\cite{Kop98}, and contaminant transport \cite{hel97}. The predictions
are obtained either from approximate theories which link the physical
properties to geometrical observables, from geometrical models for
which the physical problem can be solved more easily, or from various 
cross property relations, which relate the parameter in
question to other physical transport parameters.

In this context, the exact numerical calculation of transport properties
serves three purposes: (i) testing and validation of theories and
theoretical predictions, (ii) comparison of geometrical models, and
(iii)  testing of faithfulness of computerized tomographic imaging by
comparing numerically calculated transport parameters with their
experimental values.

Of particular interest for porous media is the permeability, and more
precisely its fluctuations. These fluctuations are important because
they dominate the large scale permeability.  For this reason it is
important to collect as many micropermeametry measurements as possible. 
Experimental micropermeametry is costly and inaccurate. Hence exact
numerical calculations are becoming an interesting alternative for
studying fluctuations in permeability. 

In this article we compare the permeabilities of a three-dimensional
computerized tomographic image of Fontainebleau sandstone and its three
physical and stochastic reconstruction models. We find that
especially the stochastic models fail in reconstructing the fluid
permeability of the original sandstone. This finding is in good
agreement with conclusions drawn previously from a purely geometrical
characterization of the same microstructures \cite{hil99c}. 

The exact numerical calculation of transport parameters for digitized
three-dimensional samples remains a computationally demanding task, and
only few studies exist which test the accuracy of such calculations. Here,
we compare the results obtained by means of a finite difference (\FD) 
method and a lattice-Boltzmann (\LB) algorithm.  We begin our study
by calibrating both simulation methods against exact solutions of the
Stokes equation for straight tubes and cubic arrays of spheres. 
These calculations also serve to compare the speed of both methods.
We then proceed to apply both algorithms to the experimental sample
and its three models.

\section{Definition of the problem}
The problem to be solved is that of slow laminar flow through a
three-dimensional porous medium on a microscopic level. The
three-dimensional microstructure of a two-phase porous medium $\SSS$
consisting of a pore phase $\PPP$ and a matrix or rock phase $\MMM$
with $\SSS = \PPP \cup \MMM$ is described in detail by the
characteristic function $\ch_\GGG$ of a single phase
$\GGG\in\{\MMM,\PPP\}$ with
\begin{equation}
  \ch_\GGG(\vec x) = \left\{\begin{array}{ll}
        1 & \hbox{for}\quad \vec x \in \GGG \\
        0 & \hbox{for}\quad \vec x \not\in \GGG 
      \end{array}\right.
  \quad .
\end{equation}
In the following $\vec x$ is the position vector of a cubic lattice,
$\vec x=a\cdot x_1\vec e_1 + a\cdot x_2\vec e_2 +a \cdot x_3\vec e_3$,
with $x_i=0,1,\dots, M_i-1$, the unit vectors $\vec e_i$ of the
Cartesian coordinate system, and the grid spacing $a$. The total
number of lattice points is given by $N=M_1M_2M_3$. 

The Reynolds numbers of interest in geophysical and petrophysical
applications are usually much smaller than unity \cite{hil94c} and hence
it suffices to solve the Stokes equation. In the pore space geometry
described by the characteristic function $\ch_\PPP(\vec x)$, the
steady-state Stokes equation and the constraint of incompressibility
read
\begin{eqnarray}
  \eta \Delta\vec v(\vec x) - \vec \nabla p(\vec x) &=& 0; \quad \vec x \in \PPP
  \label{eq:stokes},\\
  \vec \nabla \cdot \vec v(\vec x) &=& 0; \quad \vec x \in \PPP \label{eq:div0}
  \quad .
\end{eqnarray}
On the pore-matrix interface $\partial \PPP$ we apply no-flow boundary
conditions,
\begin{equation}\label{eq:bc}
  \vec v(\vec x) = 0; \quad \vec x \in \partial\PPP
  \quad .
\end{equation}
Darcy's law permits us to compute the $j$-th column of the macroscopic
permeability tensor from the microscopic solution of the hydrodynamic
problem Eqs. (\ref{eq:stokes}) - (\ref{eq:bc}) for a pressure gradient
applied along the $\vec e_j$ direction according to
\begin{equation} \label{eq:darcy}
  k_{ij}=\bra v_i(\vec x)\ket_{\vec x\in \SSS} 
  \frac{\eta (M_j+1) a}{p_{\rm in}-p_{\rm out}}
  \quad .
\end{equation}
Here the pressures at the inlet and the outlet surface are given by
$p_{\rm in}$ and $p_{\rm out}$ as defined below in Eqs. (\ref{eq:pout})
and (\ref{eq:pin}) \footnote{Due to the spatial discretization
  $p_{\rm in}$ and $p_{\rm out}$ are the pressure values at a distance
  $a/2$ from the sample surface. Hence, we have to divide by
  $(M_j+1)a$ to obtain the mean pressure gradient.} , respectively,
$\bra\dots\ket_{\vec x \in \SSS}$ denotes an average over all lattice
points, and $v_i=\vec v\cdot \vec e_i$.

\section{Samples}

We investigate four different samples $\SSS_{\EX}$, $\SSS_{\DM}$,
$\SSS_{\GF}$, and $\SSS_{\SA}$. The first sample $\SSS_{\EX}$, or in
abbreviated form \EX, was obtained experimentally by means of computerized
tomography from a core of Fontainebleau sandstone. This sandstone is a
popular reference standard because of its exceptional chemical,
crystallographic and microstructural simplicity \cite{BZ85,BCZ87}.
Fontainebleau sandstone consists of crystalline quartz grains that
have been eroded for long periods before being deposited in dunes
along shore lines during the Oligocene, i.e. roughly 30 million years
ago.  It is well sorted containing grains of around $200$ $\mu$m in
diameter.  During its geological evolution, which is still not fully
understood, the sand was cemented by silica crystallizing around the
grains.  Fontainebleau sandstone exhibits intergranular porosity
ranging from $0.03$ to roughly $0.3$ \cite{BCZ87}. The computer-assisted 
microtomography was carried out on a micro-plug drilled from
a larger original core. The original core from which the micro-plug
was taken had porosity $\phi^* = 0.1484$, permeability $k^* =
1.3$ D, ($1$ D=0.987 $\mu$m$^2$) and formation factor $22.1$ (dimensionless
electrical resistivity \cite{dul92}). 
The microtomographic dataset has dimension
$M_1\times M_2\times M_3=299\times 300 \times 300$ with a resolution
of $a=7.5\mum$, and porosity $\phi=0.1355$. The pore space
$\PPP_{\EX}$ is visualized in Fig. 1 of Ref. \cite{hil99c}. 

The three remaining samples are physical and stochastic reconstruction
models for the Fontainebleau sample \EX. All have the same lattice
resolution, $a=7.5 \mu$m,  and approximately the same porosity.  
The ``diagenesis model'' \DM tries to mimic the geological 
formation process of the
natural sandstone in three steps: the sedimentation of spherical
grains, the compaction of the sediment, and the simulation of quartz
cement overgrowth. The sample dimensions are $M_1\times M_2\times
M_3=255\times 255 \times 255$.

The \SA and \GF samples are stochastic models with dimensions 
$M_1\times M_2\times M_3=256\times 256 \times 256$. Both models
reconstruct the porosity $\phi$ and the two-point correlation function
of the original sandstone \EX. This implies the reconstruction of the
specific surface $S_V$. However, due to problems in the reconstruction
procedure of the Gaussian field method, the porosity and the specific
surface of the \GF model do not match exactly those of the
original sandstone.  We find $\phi=0.1354$ for \SA, and $\phi=0.1421$
for \GF.

For a more detailed description of the modeling procedures and a
visualization of the microstructures, the reader is referred to
Ref. \cite{hil99c} and the references therein. In the same article 
results of an extensive geometrical investigation of the four
samples are presented, which use both 
classical geometric quantities but also concepts
introduced in local porosity theory \cite{hil91d,hil96}. 
The main findings in Ref. \cite{hil99c} were as follows:
\begin{enumerate}
\item None of the three models can reproduce the visual appearance of
  the original sandstone. \EX shows a granular structure of the
  matrix phase where single sand grains can easily be identified. The
  matrix phase of \DM is also clearly granular but with
  artificial, spherical grains. In both models the matrix and the
  pore phase are very well connected, and the pore-matrix interface is
  smooth. In contrast, the pore-matrix interface of the stochastic
  models \SA and \GF is very rough. Here, both phases are strongly
  scattered and exhibit isolated clusters.
  
\item The two-point correlation functions, $\corfct(r)=\bra\ch(\vec
  x_1)\ch(\vec x_2) \ket$ with $r=|\vec x_1-\vec x_2|$, of \SA and
  \GF show good agreement with the original sandstone except for
  minor deviations at small $r$ in the case of \GF. The correlation
  function of \DM clearly deviates from that of \EX. Moreover, it shows
  strong anisotropy with respect to directions $\vec e_1$, $\vec
  e_2$ and $\vec e_3$.
  A modified correlation function giving the conditional probability
to find two points in pore space that are also connected by a path
inside the pore space was measured and discussed in Ref.\cite{hil00f}.
Small but significant differences exist between the samples.
The experimental sample \EX is more stable under the
morphological operations of erosion and dilation \cite{hil00f,hil00}.

\item The differences between the samples are most pronounced 
  when comparing the geometrical
  connectivity of the pore space. As a measure for the geometrical
  connectivity, we use the total fraction of percolating cells $p_3(L)$
  at scale $L$, introduced in local porosity theory \cite{hil00}, which
  is defined as the probability for a cubic subblock of size $L$ of
  the sample to percolate in all three coordinate directions $\vec e_i$.
  Here, percolation in direction $\vec e_i$ means
  that there exists a path lying entirely in the pore space, which spans from
  one face of the cubic subsample perpendicular to $\vec e_i$ to the
  opposite face.  For \EX and \DM the curves of $p_3(L)$ nearly
  coincide while for the stochastic models \SA and \GF the curves of
  $p_3(L)$ fall well below that of the original sandstone (see Fig. 13
  in Ref. \cite{hil99c}). Table \ref{tab:geo} gives the values of $p_3(L)$
  for $L=60a$. Again we find anisotropy in the \DM model when we
  measure the probability $p_{\vec e_i}(L)$ for a subblock to
  percolate in directions $\vec e_i$.
  
\end{enumerate}
Geometrical connectivity is an indispensable precondition for
dynamical connectivity and physical transport. Hence, we expect to
find a strong correlation between the total fraction of percolating
cells $p_3(L)$ and the macroscopic permeability whose calculation is
discussed next.

\section{The numerical methods}
\subsection{The finite difference method}
\subsubsection{Algorithm}

Numerically we obtain the solution of Eqs. (\ref{eq:stokes}) -
(\ref{eq:bc}) from the infinite time limit of the time dependent
Stokes problem using an iterative pressure-correction algorithm
\cite{B:pat80,B:hir88}. Discretization in time of the
time-dependent Stokes equation yields
\begin{eqnarray}
  \frac{\vec v^{n+1} (\vec x) - \vec v^n(\vec x)}{\Delta t} &=&
  \eta \Delta\vec v^n(\vec x) - \vec\nabla p^{n+1}(\vec x), 
  \label{eq:stokesd}\\
  \vec\nabla \cdot \vec v^{n+1}(\vec x) &=& 0, \label{eq:div0d}
  \quad 
\end{eqnarray}
where the superscript $n$ denotes the iteration step. In our case the
discretized time derivative on the left hand side of
Eq. (\ref{eq:stokesd}) has no physical meaning. In the long-time limit
$\vec v^{n+1} (\vec x) = \vec v^n(\vec x)$ holds, and we recover
Eq. (\ref{eq:stokes}).

Given the solutions $\vec v^n$ and $p^n$ at iteration step $n$, an
approximate solution $\vec v^\star$ for the velocity field is
obtained from 
\begin{equation}\label{eq:vapprox}
  \frac{\vec v^\star (\vec x) - \vec v^n(\vec x)}{\Delta t} =
  \eta \Delta\vec v^n(\vec x) - \vec\nabla p^n(\vec x) 
  \quad .
\end{equation}
Subtracting Eq. (\ref{eq:vapprox}) from Eq. (\ref{eq:stokesd}), 
we find that
\begin{equation}\label{eq:vsub}
  \frac{\vec v^{n+1} (\vec x) - \vec v^\star(\vec x)}{\Delta t} =
  - \vec\nabla \Big(p^{n+1}(\vec x) - p^n(\vec x) \Big)
  \quad .
\end{equation}
On the right hand side of this equation appears the 
pressure correction $p^\prime(\vec x) \equiv p^{n+1}(\vec x) - p^n(\vec x)$.
Applying the $\vec\nabla$ operator to Eq. (\ref{eq:vsub}), and using the
incompressibility constraint Eq. (\ref{eq:div0d}) we obtain Poisson's
equation for $p^\prime$,
\begin{equation}\label{eq:pprime}
  \Delta p^\prime(\vec x) = \frac{1}{\Delta t}\vec \nabla \cdot \vec v^\star(\vec x)
  \quad .
\end{equation}
Thus, we arrive at the following algorithm:
\begin{enumerate}
\item Let $\vec v^n$ and $p^n$ be the solution of the velocity and the
  pressure field, respectively, at iteration step $n$ with the
  maximum absolute error $\epsilon^n = \max_{\vec x\in\PPP}|\eta\Delta
  \vec v^n(\vec x)-\nabla p(\vec x)|$. From $\vec v^n$ and $p^n$ an
  approximate solution $\vec v^\star$ of the velocity field is
  calculated using Eq. (\ref{eq:vapprox}).
  
\item Using the definition of the pressure correction, a new pressure
  $p^{n+1}=p^n+p^\prime$ is obtained from a solution of
  Eq. (\ref{eq:pprime}). This part of the algorithm consumes most of
  the computation time, because Eq. (\ref{eq:pprime}) has to be solved
  for each iteration step. However, we found that it suffices to solve
  Eq. (\ref{eq:pprime}) only up to an error
\[ \max_{\vec x\in \PPP} 
| \Delta p^\prime(\vec x)-
\frac{1}{\Delta t} \vec \nabla \cdot \vec v^\star(\vec x) | 
\le \gamma \epsilon^n
\]
 where $\gamma$ is an empirical factor. For 
  the calculations presented here, a value of $\gamma$ in the range
  $0.01 < \gamma < 0.1$ seems to be appropriate.
  
  We use a successive over-relaxation method to solve
  Eq. (\ref{eq:pprime}). Of course it would be desirable to use more
  sophisticated methods such as {\em e.g.} a multigrid method, but we could not
  find a general procedure to restrict the microstructure of the
  porous medium to a coarser grid without changing the topology of the
  pore space.
  
\item From $p^\prime$ and $\vec v^\star$ a new velocity $\vec
  v^{n+1}$ is calculated using Eq. (\ref{eq:vsub}). The algorithm
  terminates when $\epsilon^{n+1}$ is smaller than some given value.
\end{enumerate}

\begin{figure}
  \psfrag{p1}[c][c]{\small$p(\vec x)$}
  \psfrag{p2}[c][c]{\small$p(\vec x+a\vec e_\perp)$}
  \psfrag{vx1} {\small$v_\perp(\vec x)$}
  \psfrag{vy1}[c][c] {\small$v_\parallel(\vec x)$}
  \psfrag{vx2} {\small$v_\perp(\vec x+a\vec e_\perp)$}
  \psfrag{vy2}[c][c] {\small$v_\parallel(\vec x+a\vec e_\perp)$}
  \psfrag{a}[c][c]    {\small$a$}
  \psfrag{ex}   {\small$\vec e_\perp$}
  \psfrag{ey}   {\small$\vec e_\parallel$}
  \psfrag{dp}   {\small$\partial \PPP$}

  \begin{center}
    \epsfig{file=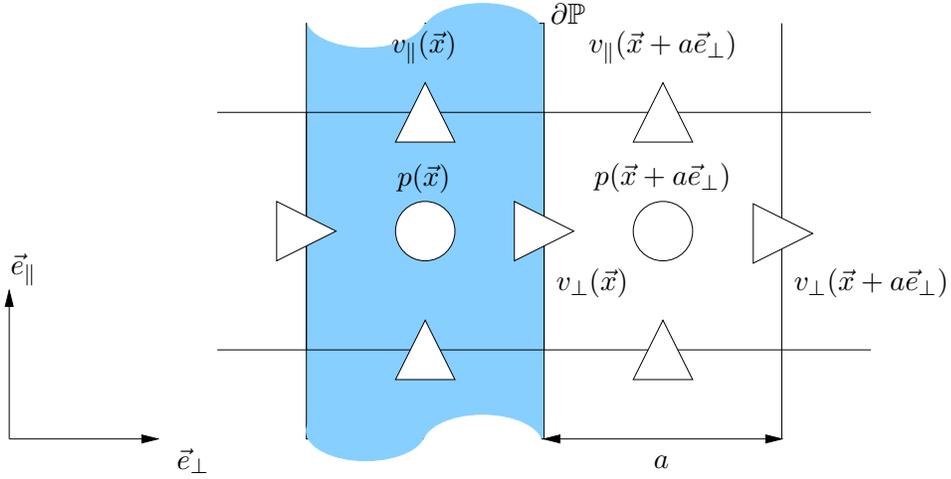,width=.7\linewidth}
  \end{center}
  
  \caption{Spatial discretization of the velocity and pressure field
    on a MAC grid. The grey shaded cell lies in matrix phase, the
    white cell in pore space.}
  \label{fig:1}
\end{figure}

The equations are spatially discretized using a marker-and-cell (MAC)
grid \cite{HW65}. The pressure values are placed at the centers of the
grid cells. They coincide with the lattice points of the discretized
characteristic function $\ch_\PPP$. On each face of a grid cell the
velocity component perpendicular to this face is located. The
pore-matrix interface $\partial \PPP$ follows the surface of the cubic
grid cells. For velocity components perpendicular to the interface
the boundary condition Eq. (\ref{eq:bc}),
\begin{equation}\label{eq:vperp-bc}
  v_\perp(\vec x) = 0,
\end{equation}
is implemented exactly. For parallel velocity components at 
distance $a/2$ from the interface, Eq. (\ref{eq:bc}) is fulfilled
to second order accuracy,
\begin{equation}\label{eq:vpar-bc}
  v_\parallel (\vec x) = -v_\parallel(\vec x+a\vec e_\perp) + O(a^2),
\end{equation}
in which the interface is located as shown in Fig. \ref{fig:1}. Inserting
Eq. (\ref{eq:bc}) into the Stokes equation Eq. (\ref{eq:stokes}) one finds as
the boundary conditions for the pressure
\begin{equation}\label{eq:p-bc}
  \left.\partial_\perp p(\vec x) \right|_{\partial \PPP} = 0
  \quad .
\end{equation}
Using Eqs. (\ref{eq:vperp-bc}) - (\ref{eq:p-bc}), the $\Delta$ operators
in Eqs. (\ref{eq:vapprox}) and (\ref{eq:pprime}) take the form
\begin{equation}\label{eq:vlap-diskret}
\begin{split}
  \Delta v_i(\vec x) &= 
    v_i(\vec x-a\vec e_i) + v_i(\vec x+a\vec e_i) - 2v_i(\vec x)\\
    &+ \sum_{j\ne i}\bigg[
      \ch_\PPP(\vec x+a\vec e_j)\ch_\PPP(\vec x+a\vec e_j+a\vec
      e_i)\big[v_i(\vec x+a\vec e_j)+v_i(\vec x)\big]\\ 
    &+ \ch_\PPP(\vec x-a\vec e_j)\ch_\PPP(\vec x-a\vec e_j+a\vec
      e_i)\big[v_i(\vec x-a\vec e_j)+v_i(\vec x)\big]
      - 4 v_i(\vec x)\bigg] \quad ; \vec x \in \PPP ,
\end{split}
\end{equation}
and
\begin{equation} \label{eq:plap-diskret}
  \begin{split}
  \Delta p^\prime(\vec x) = \sum_i\bigg[
  &\ch_\PPP(\vec x+a\vec e_i)\big(p^\prime(\vec x+a\vec e_i) -
  p^\prime(\vec x)\big)\\
  &\ch_\PPP(\vec x-a\vec e_i)\big(p^\prime(\vec x-a\vec e_i) -
  p^\prime(\vec x)\big)
  \bigg] \quad ; \vec x \in \PPP ,
  \end{split}
\end{equation}
where we used second order accurate central differences to discretize
the spatial derivatives.

\subsubsection{Boundary conditions}

On the sample surface an additional outer layer of grid cells, the
so-called shadow row, is added. It provides the neighboring pressure
and velocity values needed for the evaluation of
Eqs. (\ref{eq:vlap-diskret}) and (\ref{eq:plap-diskret}) for the grid cells
on the sample surface. The pressure and velocity values of the shadow
row are set according to the macroscopic boundary conditions. Let
$\vec e_j$ denote the direction of the applied pressure gradient.  For
grid cells of the shadow row on the outflow boundary of the sample,
i.e.  $\vec x \in \{\vec x : x_j=M_j\}$, we choose
\begin{eqnarray} \label{eq:bcout}
  p(\vec x) &=& p_{\rm out} ,\label{eq:pout}\\
  v_i(\vec x) &=& 0 \quad \hbox{for\ } i\ne j ,\\
  v_j(\vec x) &=& v_j(\vec x-a\vec e_\parallel) ,
\end{eqnarray}
as the boundary conditions. For $\vec x \in \{\vec x : x_j=-1\}$, i.e. on
the inflow boundary, we set
\begin{eqnarray}
  p(\vec x) &=& p_{\rm in} ,\label{eq:pin}\\
  v_i(\vec x) &=& 0 \quad \hbox{for\ } i\ne j ,\\
  v_j(\vec x) &=& v_j(\vec x-a\vec e_\parallel) ,\label{eq:bcin}
\end{eqnarray}
as the boundary conditions. The last condition has to be introduced
directly into Eq. (\ref{eq:vlap-diskret}) because the position $\vec
x-a\vec e_j$ lies outside the simulation lattice.
In the simulations reported below we used always $p_{\rm in}=1$,
$p_{\rm out}=-1$ unless indicated otherwise.

On the remaining sample surfaces the grid cells of the shadow row are
assigned to the matrix phase. The velocity and pressure values are set
to zero.

\subsection{ The lattice-Boltzmann method}
In this section we introduce the lattice-Boltzmann method used here,
in particular the LBGK (lattice-Bhatnagar-Gross-Krook) model. We
discuss then the basic hydrodynamics of the model and the relevant
boundary conditions. The numerical accuracy of the lattice-Boltzmann
results for permeability will be considered in terms of finite-size
effects. 
\subsubsection{Lattice-Boltzmann hydrodynamics}

The lattice-Boltzmann method \cite{Qia92,Ben92,Rot97,CD98} is a
mesoscopic approach for computational fluid dynamics in which the
basic idea is to solve a discretized Boltzmann equation.  The
macroscopic dynamics of the system can be shown to obey the
Navier-Stokes equation.  One of the most successful applications of
the method has been to flow in porous media
\cite{Fer95,Mar96,Kop98}. 

In this method the fluid is modeled by particle distributions that move
on a regular lattice. In our implementation each lattice point is
connected to its nearest and next nearest neighbors. Together with a
rest particle, each lattice point is then occupied by 19 different
particles (the D3Q19 model). At each time step particles propagate to
their adjacent lattice points, and re-distribute their momenta in the
subsequent collisions. The dynamics of the LBGK model is given by the
equation \cite{Qia92,Ben92}
\begin{equation}
f_i({\bf r}+{\bf c}_{i},t+1)=f_i({\bf r},t)+\frac{1}{\tau}
[f_i^{\rm eq}({\bf r},t)-f_i({\bf r},t)],\label{bgk}
\end{equation}
where ${\bf c}_{i}$ is a vector pointing to an adjacent lattice site,
$f_i({\bf r},t)$ is the density of the particles moving in the ${\bf
  c}_{i}$-direction, ${\tau}$ is the BGK relaxation parameter, and
$f_i^{\rm eq}({\bf r},t)$ is the equilibrium distribution towards
which the particle populations are relaxed. Hydrodynamic quantities
like density $\rho$ and velocity ${\bf u}$ are obtained from the
velocity moments of the distribution $f_i$
in analogy with the kinetic theory of gases. The equilibrium
distribution can be chosen in many ways. A common choice is
\begin{equation}
f_i^{\rm eq}=t_i(1+\frac{1}{c_s^2}({\bf c}_{i}\cdot 
{\bf u})+\frac{1}{2c_s^4}({\bf c}_{i}\cdot {\bf u})^2-\frac{1}{2c_s^2}u^2),
 \label{eq}
\end{equation}
in which $t_i$ is a weight factor that depends on the length of the link
vector ${\bf c}_{i}$, and $c_s$ is the speed of sound in the fluid.
The weights $t_i$ we choose here in accordance with the 19-link LBGK
model, and they are $\frac{1}{3},\frac{1}{18}$ and $\frac{1}{36}$ for
the rest particle and the particles moving to the nearest and
next-nearest neighbor sites, respectively. The speed of sound is
$c_s=\frac{1}{\sqrt 3}$ for this model, and the kinematic viscosity of
the simulated fluid is ${\eta}=\frac{2{\tau}-1}{6}$.
(Here and in the following, lattice units are always used if the units
are not specified.) The fluid pressure is given by
\begin{equation}
p({\bf r},t)=c_s^{2}(\rho({\bf r},t)-\bar \rho)\equiv
c_s^{2}\Delta\rho({\bf r},t),
\end{equation}
where $\bar \rho$ is the mean density of the fluid.

The Stokes equation, Eq. (\ref{eq:stokes}), is produced directly by
the linearized lattice-Boltzmann method, in which the quadratic velocity
terms in the equilibrium distribution function Eq. (\ref{eq}) are
neglected. To be consistent with the finite difference method, we use
in what follows the linearized lattice-Boltzmann method if not stated
otherwise.

\subsubsection{Boundary conditions}

The physical boundary condition at solid-fluid interfaces is the
no-flow condition Eq. (\ref{eq:bc}), which in lattice-Boltzmann
simulations is usually realized by the so-called bounce-back rule
\cite{Gal97,He97}. In this approach the momenta of the particles that
meet a solid wall are simply reversed. 

In simple shear flows the bounce-back condition assumes that the location
of the wall is exactly halfway between the last fluid point and the
first wall point. In more complicated cases the no-flow boundary
lies somewhere in between these two points, the exact place
depending on the relaxation parameter and the geometry of the system
\cite{He97,Kan99}. In Poiseuille flow, {\em e.g.}, the bounce-back rule 
gives velocity fields that deviate from the exact solution,
for no-flow boundaries at exactly halfway between the last fluid
point and the first solid point, by \cite{He97}
\begin{equation}
  \Delta u = u_{\rm sim} - u_{\rm exact} = u_{\rm max}\frac{48\eta^2-4\eta-1}{L^2},
  \label{tau-depen}
\end{equation}
where $u_{\rm sim}$ and $u_{\rm exact}$ are the simulated and the exact 
velocities, respectively,
$u_{\rm max}$ is the velocity at the center of the channel, and $L$ 
is channel width.
This implies that the simulated permeability will depend somewhat on
viscosity especially at low discretization levels. 
This viscosity dependence can practically be eliminated by using the 
so-called second order boundaries, in which case the desired location of the
no-flow boundary is determined by extrapolating the distribution function
from the last fluid points.
Some of these more sophisticated solid-fluid boundaries are restricted
to regular geometries \cite{Sko93,Nob95}, but there are also general
boundary-fitted models \cite{Fil98,Che98} available. For practical
simulations the bounce-back boundary is however very attractive, because it
is a simple and computationally efficient method for imposing no-flow
conditions on irregularly shaped walls. Also, the error created by
the bounce-back boundary does not destroy the spatial second-order
convergence of the method \cite{Kan99,Mai96}. 

In simulating fluid flow it is important that the velocity and
pressure boundary conditions of the system have been imposed in a
consistent way. However, general velocity and pressure boundaries are
still under development for the lattice-Boltzmann method
\cite{Mai96,Che96,Gin96}. So far in most of the practical
simulations a body force has been implemented \cite{Fer95,Kan99,Bui2000} 
instead of pressure or velocity boundaries. 

When the body force is used, the pressure gradient acting on the
fluid is replaced with a uniform external force. The use of a body
force is based on the assumption that, on the average, the effect of
an external pressure gradient is constant throughout the system, and
that it can thus be replaced with a constant force that adds at
every time step a fixed amount of momentum on the fluid points. 
Conditions that are close to pressure boundaries can be obtained
by averaging the velocity and pressure fields over the planes of the
inlet and outlet of the simulated system \cite{Kan99}. 

During one iteration step, the fluid momentum oscillates in the stationary
state by an amount given to each fluid point by the body force. For this
reason the fluid velocity is now defined as the average of the
pre-collision and post-collision values \cite{Shan93,Shan95}.


Pressure fields generated by the body force are obtained from the
effective pressure $p_{\rm eff}$,
\begin{equation}
p_{\rm eff}({\bf r},t) = c_s^2 \Delta\rho({\bf r},t) - {\bar \rho} g x,
\label{eff_press}
\end{equation}
where $x$ is the distance from the inlet of the system
measured in the flow direction, and $g$ is the acceleration
the body force gives to the fluid.

It is a well-known fact that, due to staggered invariants, the fluid
momentum may oscillate in a time scale of a few time steps \cite{CD98},
even in the stationary state. In open areas this effect is usually
unimportant, but in closed pores this effect may become visible as the
fluid momentum may oscillate around zero, with a magnitude determined by
the body force. This effect may lead to some corruption of the
fluid-velocity distributions as can be seen in Fig.\ref{fig:sand-hist}. 
Notice that staggered momenta can be eliminated by
averaging the quantities over a few time steps. 

Notice finally that the diagonal links allow the fluid to leak to
neighboring lattice points which have only a single edge in common. For
this reason the 'standard' 
lattice-Boltzmann model is not expected to be accurate
very close to the percolation threshold. 
For the 3D checkerboard
structure, {\em e.g.}, we found that the permeability of the system was
about 0.036 lattice units for all the six lattice resolutions that were
used, although the structure is not percolating.
If better accuracy is needed, diagonal leaks can be eliminated 
by applying the bounce-back rule on such diagonal links which 
actually cross a solid boundary (like diagonals in the 
checkerboard structure).

\subsubsection{Finite-size effects and the saturation time}

The accuracy of lattice-Boltzmann simulations depends on the ratio of
the mean free path $\lambda_{\rm mfp}$ of the fluid particles to the
representative size $\lambda_{\rm 0}$ of the obstacles and pores
\cite{Rot97,Fer95,Gin96}. The simulated flow field does not describe the
true hydrodynamic behavior unless this ratio is small. For increasing
$\lambda_{\rm mfp}/\lambda_{\rm 0}$ ratio {\em Knudsen-flow} behavior is
found, which is also true in real fluids \cite{Sch57}. These effects
must always be considered when lattice-Boltzmann simulations are
performed. In this way the maximum size of the lattice spacing can be
estimated together with the accuracy of the simulations.

Finite-size effects restrict to some extent the use of LB-methods based
on regular lattices. In porous media close to the percolation
threshold, {\em e.g.}, many pores are very small, and very big lattices
may be needed for realistic simulations. It is still an open question
whether the finite-size effects are always dominated by the minimum pore
size or the average pore size. 
The effect can be estimated by simulating the system with several different
lattice spacings, but occasionally it is difficult to distinguish the
finite-size effects from other sources of numerical error. 

Practice has shown \cite{Fer95,Kop98}
that smaller values of the relaxation parameter $\tau$ tend to decrease the
finite-size effects (see also our simulations below). Equation
(\ref{tau-depen}) can be used to explain this: due to Knudsen-flow
effects, low-discretization simulations regularly give too high
permeabilities, whereas decrease of $\tau$ has the opposite effect
down to $\tau=0.625$, at least for tube flows. On the other hand,
the lattice-Boltzmann algorithm may become \cite{CD98} unstable with 
values of $\tau$ close to $0.5$. In practical
permeability simulations the relaxation parameter has usually been
chosen to be bigger than 0.6.
The effect of $\tau$ on the behavior of the lattice-Boltzmann model is thus
quite complicated and not yet fully understood.

In permeability simulations a simple dimensional
analysis shows that, with a constant body force, the saturation time
$t_{\rm sat}$ needed to reach the steady state is of the form
\begin{equation}
 t_{\rm sat}\propto R_{\rm pore}^2/\eta, 
 \label{sat_time}
\end{equation}
where $R_{\rm pore}$ is the characteristic length of the void pores in
the system. For systems with high porosity, the saturation times can
therefore be very long. In some cases, tens of thousands of time steps
may be needed. It is thus evident that a constant body force may be
computationally inefficient, especially when one is only interested in
the steady-state solution. The saturation time can be reduced by using
{\it e.g.} the iterative momentum-relaxation (IMR) method, where the 
applied body force is adjusted during the iteration in a definite 
relation to the change of the fluid momentum during iteration steps 
\cite{Kan99}.
For other ways to reduce the saturation time see \cite{Ver99}.

\section{Results}
\subsection{Tube with quadratic cross section}

One of the few cases for which the analytical solution of the
hydrodynamic problem Eqs. (\ref{eq:stokes}) - (\ref{eq:bc}) is known is
Poiseuille flow, the flow through a linear tube with constant cross
section. We will consider a tube with quadratic cross section, because
here the geometry can be discretized on a cubic lattice without
discretization error. 

\begin{figure}
\begin{center}
  \epsfig{file=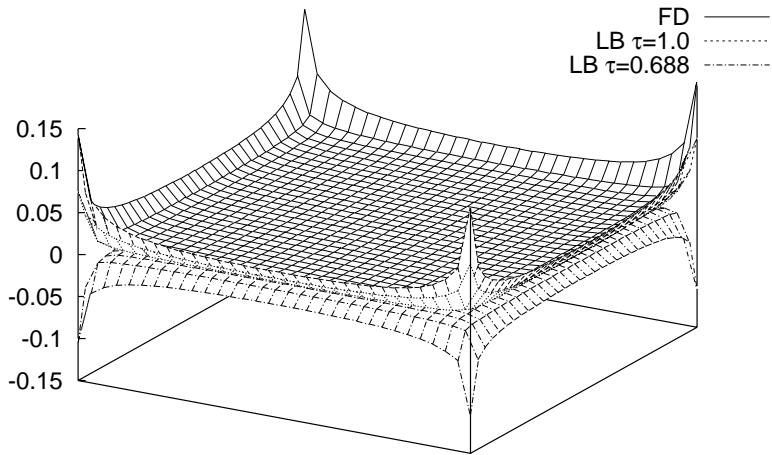, width=0.7\linewidth}
\end{center}

  \caption{Relative error of the numerical solution of the velocity field 
    for a Poiseuille flow through a tube with quadratic cross section
    of size $B=32a$. The upper surface shows the \FD solution, the
    lower surfaces the \LB solutions with $\tau=1.0$ and $\tau=0.688$,
    respectively. As the \FD solution, the \LB solution with $\tau=1.0$
    overestimates the reference solution while the \LB solution with
    $\tau=0.688$ underestimates the references values. The reference
    values are calculated from the analytical solution given in
    Ref. \cite{B:wie74}.}
  \label{fig:channel}
\end{figure}

We consider a tube directed along the $\vec e_1$ direction with
quadratic cross section of side length $B=32a$. We compare the velocity
component $v_1(x_2, x_3)$ with its exactly known reference value $v^{\rm
ref}_1(x_2, x_3)$ given in Ref. \cite{B:wie74}. Figure \ref{fig:channel}a
shows the relative error $(v_1-v_1^{\rm ref})/v_1^{\rm ref}$ for the \LB
solution with relaxation parameter $\tau=0.688$. In Fig.
\ref{fig:channel}b we show the same relative error for the \LB solution
with $\tau=1.0$ (lower surface) and the \FD solution (upper surface).
The pressure gradient in the \FD simulation was $2/33$ while
in the \LB simulations it was around $10^{-4}$.

Around the center of the tube the analytical flow profile is very well
recovered. Near the boundaries we find deviations which are biggest in
the corners. The \LB solution with $\tau=0.688$
underestimates the reference value $v^{\rm ref}$ while the solution with
$\tau=1.0$ overestimates the true value. Hence, the relaxation parameter
$\tau$ or equivalently the viscosity $\eta$ could be adjusted
to find better agreement with the analytical velocity field.  From Fig. 
\ref{fig:channel} one expects to find a value $0.688 < \tau < 1.0$ for
which the numerical solution closely matches the analytical velocity
profile.

The computation time needed by the \FD method which terminated when
$\max_{\vec x \in \PPP}|\Delta \vec v(\vec x)-\nabla p(\vec x)|<10^{-8}$
was 951 s on a DEC Alpha workstation. In \LB simulations, the
relative error of permeability was below $10^{-5}$ in 754 s on 
a Cray T3E for $\tau=1.0$, but the simulations were continued for over 
5000 s to make sure the saturation of the velocity fields.

\subsection{Cubic array of spheres}

To test the accuracy and efficiency of our two algorithms in a more
complicated geometry with narrow constrictions we computed 
flow past a cubic array of spheres. This problem has become a
reference system for checking hydrodynamic algorithms because
accurate reference values for the permeability, and the drag
coefficient, are available over a wide range of porosities
\cite{LH89,lad88}. 

The solution of this problem proceeds by solving the problem in a
single unit cell of the cubic lattice. We generated six different
unit cells of size $L \in \{20a,36a,56a,63a,71a,89a\}$. A sphere is
placed at the center of each cell whose radius is chosen such that
the porosity matches as close as possible to $\phi=0.15$. Thus, the
porosity is close to the porosity of the sandstones investigated
later.

The \FD solution of the flow field was computed using periodic
boundary conditions on those faces of the unit cell that are
parallel to the macroscopic flow direction.
On the faces of the unit cell perpendicular to the macroscopic flow 
we applied the conditions Eqs. (\ref{eq:bcout}) - (\ref{eq:bcin})
with the standard pressure gradient $p_{\rm in}=1$, $p_{\rm out}=-1$.
In the \LB solution the flow field was computed using periodic
boundary conditions in all directions. 
This difference of the boundary conditions arises from the fact 
that in the \LB simulations 
the density fluctuations around an average density are calculated
while in the \FD simulations the pressure field enters directly.

Once the velocity field was known we calculated the permeability from 
Eq. (\ref{eq:darcy}).
Following \cite{LH89} we then utilize the expression 
\begin{equation}
  \frac{k}{R^2}=\frac{1}{6\pi C_D}\left(\frac{L}{R}\right)^3 ,
\end{equation}
to obtain the reference value $k_{\rm ref}$ of the permeability
from the drag coefficient $C_D$
given in Ref. \cite{LH89}.
The radius $R(\phi,L)$ of the spheres depends on $\phi$ and $L$ and is given 
implicitly by the expression
\begin{equation}
  \phi = \frac{8\pi}{3}\left(\frac{R}{L}\right)^3 - 3\pi
  \left(\frac{R}{L}\right)^2 + \frac{\pi}{4} + 1
  \quad .
\end{equation}
To calculate the reference value $k_{\rm ref}$ we solve this equation
and find $(L/R)\approx 1.6011$ for $\phi=0.15$. Using the drag
coefficient $C_D=1.020\times 10^3$ \cite{LH89}, we find $k_{\rm
  ref}/R^2\approx 0.0002135$.

\begin{figure}
  \begin{center}
    \epsfig{file=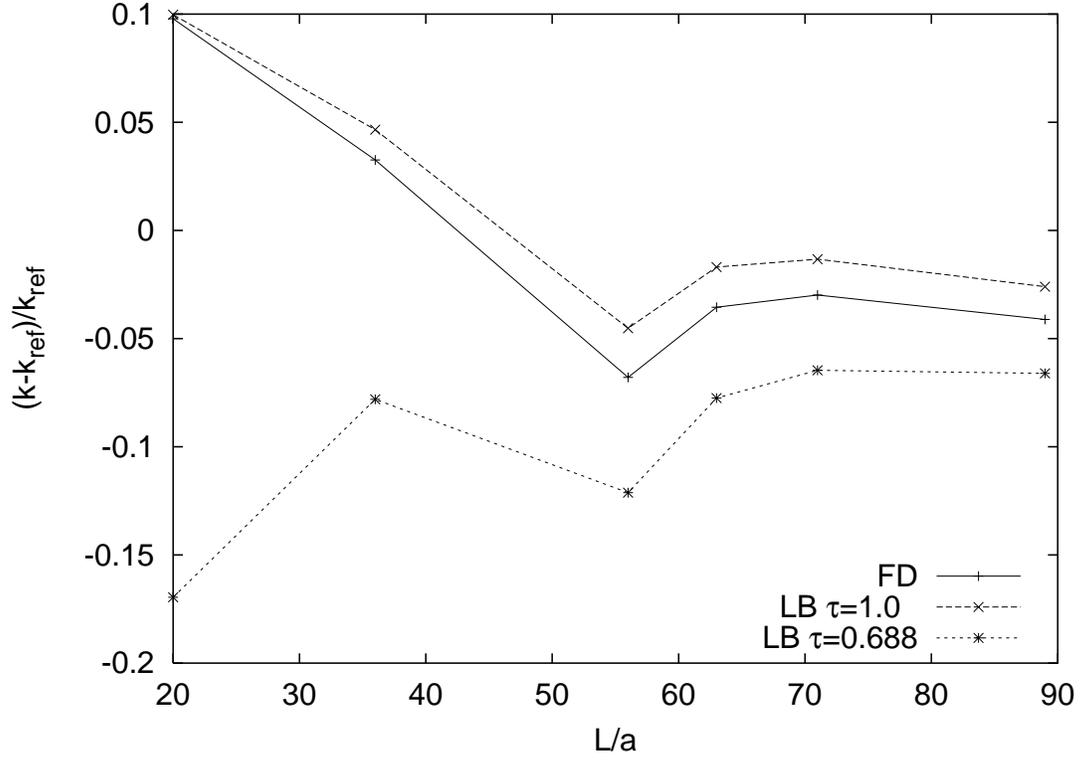, width=0.9\linewidth}
  \end{center}

  \caption{Numerical results for the permeability $k$ of a cubic array of
    spheres for different values of the lattice spacing $a$. The
    porosity is constant for all systems, $\phi=0.15$. The reference
    value $k_{\rm ref}$ is taken from Ref. \cite{LH89}.}
  \label{fig:k-a}
\end{figure}

In Fig. \ref{fig:k-a} the relative error $(k-k_{\rm ref})/k_{\rm
ref}$ of the permeability is plotted as a function of the linear 
dimensionless system size $L/a$. With increasing resolution the
results of both methods converge to each other and the error
predominantly decreases. In the lattice-Boltzmann simulations the 
relaxation parameter $\tau=1.0$ is seen to give regularly better
results than $\tau=0.688$. It thus appears that for $\tau=1.0$ the 
effective location of the no-flow boundary is more satisfactory than 
for $\tau=0.688$, and, consequently, the relative error is smaller 
for $\tau=1.0$ even though the finite-size effects are similar in
both cases. The deviation for $L=89a$ is negative and varies between 
$3$\% and $6$\% depending on the method.
This discrepancy might possibly result mainly from discretization errors
as there is a similar oscillatory trend in the \LB as well as the \FD
results. Further work on larger systems is however needed to answer
the question whether the discrepancy might also result from other sources.

The curves in Fig. 3 are all nonmonotonous. This results most
likely from the discretization of the cross-sectional area of the
pore throats between the spheres. The curve of the discretized
cross-sectional area as a function of $L/a$ shows a similar nonmonotonic
behavior.

Besides the accuracy, the demand of computation time is the second
important characteristic of a numerical method. Comparison of the
computation time of the \LB and the \FD algorithms is difficult. 
The \FD method iterates the physical velocity field $\vec v$ and the
physical pressure field $p$. The iteration is terminated when $\vec
v$ and $p$ fulfill the Stokes equation Eq. \eqref{eq:stokes} with a
predefined accuracy. In the \LB method on the other hand, the
particle distributions $f_i$ are iterated. The velocity and pressure fields 
are calculated from the final density distribution. In our
implementation there is no deterministic stopping criterion, 
although such criterion 
could be included. In practice the calculations were terminated after
a fixed number of iteration steps. This number was determined for
each system by comparing runs of different length. 

All calculations with the \FD code were performed on a Cray
T3E-900/512 at the HLRS computing center of the University of
Stuttgart with a peak performance of 461 Gflops. The \LB code was run
on a Cray T3E-750/512 at Center for Scientific Computing (CSC) in
Espoo, Finland, with a peak performance of 384 Gflops. In order to
compare the run times for the two codes we took the actual time
required to execute the program corrected by the ratio of peak
performances.
The run time required for the \LB code was calculated
from the number of iterations multiplied by a conversion factor.  
The conversion factor was  0.039448  for $L=20a$, 
0.208096 for $L=36a$, 0.628828 for $L=56a$, 0.885556 for
$L=63a$, 1.275176 for $L=71a$, and 2.21636 for $L=89a$.
It was determined by the wall time spent for one iteration
step computed from averages over several 100-step iterations.
The memory requirements of the two algorithms are different.
The \FD algorithm requires to store 8 numbers per lattice
node in the version used here. 
The D3Q19 model used for the \LB-algorithm requires to store 19
numbers per lattice node 

\begin{figure}
\begin{center}
  \epsfig{file=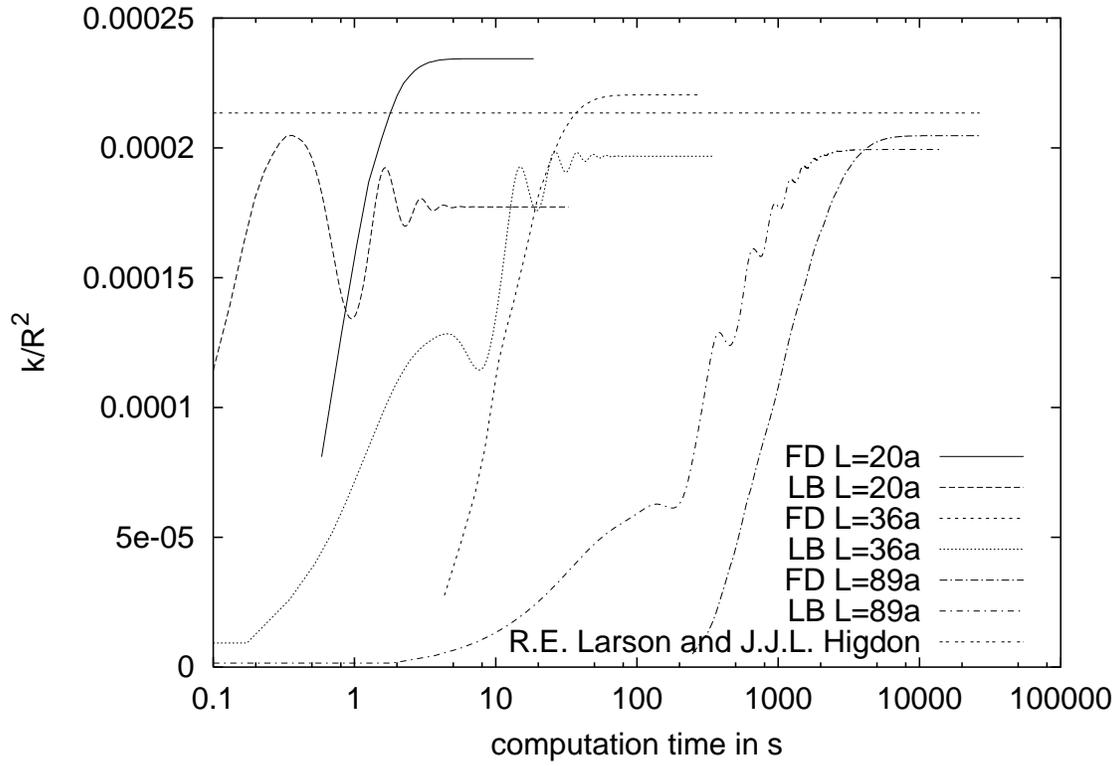, width=0.9\linewidth}
\end{center}

  \caption{Time evolution of the numerical solution of the
    permeability $k$ for flow through a cubic array of spheres with
    porosity $\phi=0.15$. The relaxation parameter was $\tau=0.688$ 
    for the \LB simulations.}
  \label{fig:k-t}
\end{figure}

In Fig. \ref{fig:k-t} we compare the time evolution of the numerical
value of the permeability $k$ for different system sizes. Plotted on
the $x$ axis is the total time in seconds needed on two
Cray-processors.  For $L=20a$ and $L=36a$, both methods reach the
final value of $k$ in approximately the same time. For large $L$
the \LB method seems to be faster. 
Notice that the results of the \LB simulations
shown in Fig. \ref{fig:k-t} were performed for $\tau=0.688$. For $\tau=1.0$ 
the simulations were about $45$ \% faster.

The convergence of $k(t)$ towards its asymptotic value is monotonic for
the \FD method, while in the \LB case $k(t)$ shows strong oscillations
for all $L$. The reason for these oscillations is probably the slight
compressibility error inherent in the model \cite{Rot97,CD98}. The horizontal line in
Fig. \ref{fig:k-t} gives the reference value $k_{\rm ref}$ towards which
the asymptotic values of both algorithms converge with $L$. 

We also compared the permeabilities given by the Navier-Stokes and
Stokes (linearized) versions of the \LB method for the cubic arrays of
spheres. Very much as expected, the two results were the same within the
first seven digits (body force about $10^{-4}$).
\subsection{Three-dimensional sandstones}

We now apply both algorithms to the solution of the hydrodynamic
problem Eqs. (\ref{eq:stokes}) - (\ref{eq:bc}) within the irregular pore
space geometries of the whole experimental sample \EX and the model
samples \DM, \GF, and \SA.
        
The \FD algorithm used the boundary conditions as described above. The
iteration scheme was terminated when the condition 
$\max_{\vec x \in \PPP}|\Delta \vec v(\vec x)-\nabla p(\vec x)|<10^{-6}$ 
for the dimensionless left hand
side of Eq. (\ref{eq:stokes}) was fulfilled for the first time. Thus, the
relative error $\epsilon(k_{ii})/k_{ii}$ of the diagonal elements of the
permeability tensor is estimated to be smaller than $0.012$ in the case
of \EX, and $\epsilon(k_{ii})/k_{ii} < 0.36$ in the case of \SA. The
relative error for the samples \DM and \GF lies in between these two
extreme values.

In the \LB simulations no flow boundary conditions were applied
on the sample surfaces parallel to the main flow direction.
At the inlet and outlet (i.e. the sample surfaces perpendicular to the 
main flow direction) an additional fluid layer with a thickness of 
$19-21$ lattice spacings was added and then periodic boundary
conditions were applied.
The body force did not act in the additional fluid layer.
These additional fluid layers increased the total number of lattice 
points by about $8$ \% in comparison with the \FD method. 
The relaxation parameter was $\tau=0.688$. 
The simulation stopped after a predefined number of iterations which was
estimated to suffice for the permeability to converge. 

In Table \ref{tab:k} we give the components of the permeability tensors
for all four samples and for both algorithms. 

The permeability results confirm the predictions from a previous purely
geometrical analysis based on local porosity theory \cite{hil99c}. The
analysis in Ref. \cite{hil99c} emphasized the importance of local
connectivity. The permeabilities are strongly correlated with the
geometrical connectivity of the pore space, measured by means of the
total fraction of percolating cells $p_3(L)$.  In accordance with our
discussion of $p_3(60a)$ given in Table \ref{tab:geo} of Section III, we
find that the permeability of the original sandstone \EX and that of the
process model \DM are in good agreement, while the permeabilities of the
stochastic models \GF and \SA are an order of magnitude smaller. It
seems as if the stochastic reconstruction models cannot reproduce the
high degree of geometrical connectivity present in the original
sandstone. The reconstructed two-point correlation function lacks
information about the geometrical connectivity of the pore space.
A correct description of the geometrical connectivity is however an
indispensable precondition for the correct dynamical connectivity that
determines the transport properties.

We now proceed to compare the numerically obtained value of the
permeability $k_{\EX}$ of the Fontainebleau sandstone \EX with the
experimental value $k^*=1.3$ D. Such a comparison requires a correction
due to the difference between the porosity of the \EX sample and the
porosity $\phi^*=0.1484$ of the original core sample on which the
experiment was performed. There exists a well-known experimental
correlation between porosity and permeability of Fontainebleau sandstone
\cite{BCZ87}. This correlation is usually approximated in the form
\begin{equation}
  k = A\phi^b ,
\end{equation}
in which $A$ and $b$ are constants. In the porosity range of interest
$\phi \approx 0.13 \dots 0.15$, this correlation has $b\approx 4$, with
however a large uncertainty due to the scatter in the measured
results. Hence, we can extrapolate the numerically determined
permeabilities $\bar k = (k_{11}+k_{22}+k_{33})/3$ into the prediction
\begin{equation}\label{eq:kstar}
  \bar k^* = \bar k \left(\frac{\phi^*}{\phi}\right)^b ,
\end{equation}
where $\phi^*$ is the previously defined porosity of the core sample and
the constant $A$ has dropped out. From Eq. (\ref{eq:kstar}) we obtain
$\bar k^* = 1150\mD$ for the \FD method and $\bar k^* = 1015\mD$ for the
\LB algorithm. These values are surprisingly close to the experimental
value $k^*=1300\mD$. Such an excellent agreement is not common. This
will in fact be seen in the following when we determine the permeability
of a subsample. 

We also checked by the \LB methods the difference between the
Navier-Stokes and Stokes permeability of sample \EX. The relative
difference was found to be $0.00036$ for the parameters specified
above, with an about $17\%$ longer simulation time in the full
Navier-Stokes case. The smallness of this difference only demonstrates
that, for the small pressure differences considered here, we indeed are
in the Stokes regime.

\subsection{Fine Graining}
We investigate the permeability of subsamples of the original samples
for two reasons. Firstly this gives us an estimate of the magnitude of
permeability fluctuations, and secondly it allows us to estimate the
dependence of numerically obtained permeability on the lattice
resolution $a$. The subsamples have dimensions of $M_1\times
M_2\times M_3 = 100\times 100 \times 100$. To test the dependence of $k$
on $a$, we apply a fine graining scheme \cite{man01}. 
The fine graining algorithm replaces each
lattice point by $n\times n\times n$ lattice points of the same phase
with $n \in \{2,3,4\}$. Thus, we get systems of dimensions $M_1\times
M_2\times M_3= n\cdot100\times n\cdot100 \times n\cdot100$ with lattice
spacing $a_n = \frac{7.5}{n}\mum$. For each system the hydrodynamic
problem Eqs.  (\ref{eq:stokes}) - (\ref{eq:bc}) is solved, and the
permeability $k^{(n)}$ is determined. 

\begin{figure}
  \begin{center}
    \epsfig{file=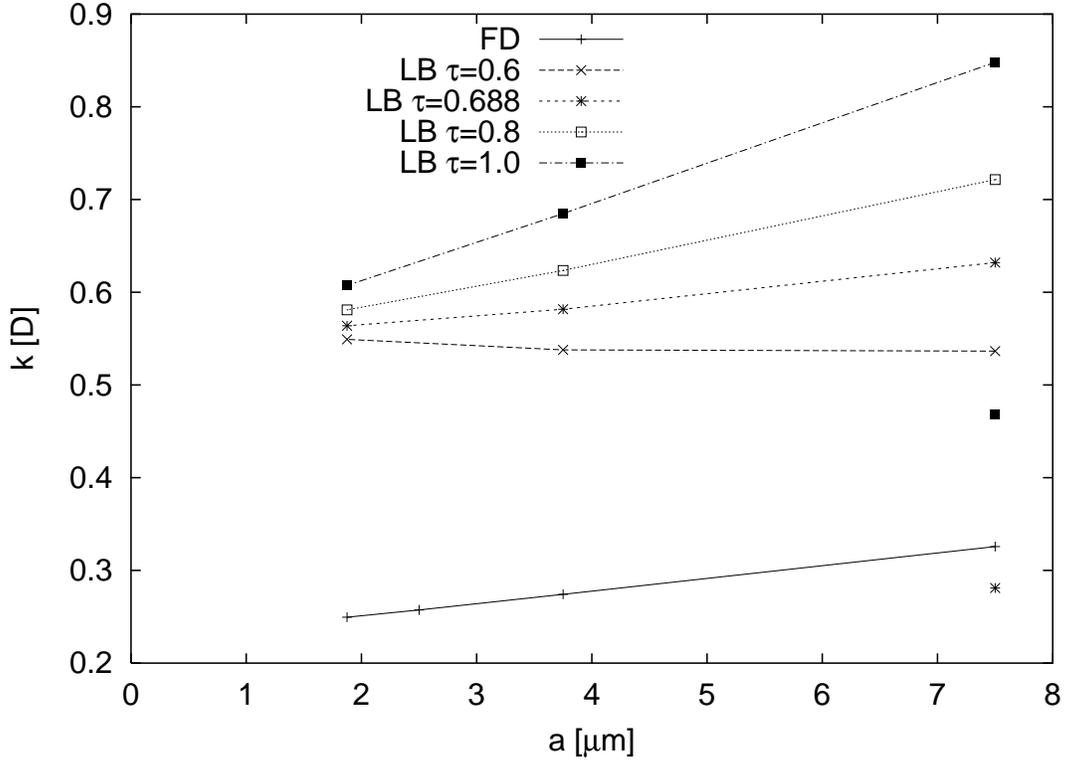, width=0.9\linewidth}
  \end{center}
  \caption{Permeability $k$ of a cubic subsample
    of the Fontainebleau sandstone (\EX) for different values of the
    lattice spacing $a$ and different relaxation parameters $\tau$.
    The size of the subsample is $L=750 \mum$. The \LB simulation used
    an additional fluid layer and the periodic boundary conditions.
    The isolated data points with
    $a=7.5$ $\mum$ are obtained from \LB simulation without extra fluid layer,
    with momentum averaging in the inlet and outlet
    and with no-flow boundary conditions. The \LB relaxations parameters
    were $\tau=1.0$ and $\tau=0.688$.}
  \label{fig:sub-k-a}
\end{figure}

Figure \ref{fig:sub-k-a} shows $k^{(n)}$ for a cubic subsample of \EX
with an applied pressure gradient in direction $\vec e_1$. The
permeabilities obtained from the \LB simulations are significantly
higher for all $\tau$. This is due to the extra fluid layer with a
thickness of $10$ \% of $M_1$ outside the sample and the periodic
boundary conditions used in the \LB simulations.  For this
smaller piece of the sample, the effect of the boundary conditions is
more significant than for the whole sample. 
We also performed with \LB a couple of simulations with conditions
similar to those used in \FD. The pressure boundary condition
was imitated using the body force combined with density and momentum
averaging at the inlet and the outlet in the adjacent free fluid 
layer with thickness of one lattice spacing. No-flow
boundary conditions were applied on the other cube sides. We found that
the results of the \LB and the \FD method were again very close. This is
indicated in Fig. \ref{fig:sub-k-a} by two isolated points (filled
square and star) at $a_1=7.5\mum$ which were obtained from \LB
simulations with these boundary conditions for $\tau=0.688$ and
$\tau=1.0$, respectively. However, a complete recomputation of the data
of Fig. \ref{fig:sub-k-a} would have exhausted the available
computation time. Moreover, the results for the free fluid layer and for
periodic boundary conditions allow us to estimate the influence of the
boundary conditions on the result. 

For the original resolution the permeabilities obtained from the \LB
algorithm differ with varying $\tau$ by nearly a factor $2$. This again
shows that the accuracy of the \LB results for low-porosity (and
low-discretization) systems strongly depends on the relaxation
parameter. While for $\tau=0.6$ the \LB results are nearly independent
of the lattice spacing, the changes in the permeability are
drastically increased with increasing $\tau$. As we have already
discussed, in the permeability simulations, decrease in the relaxation
parameter $\tau$ (or viscosity) can be used to compensate an inadequate
grid resolution (finite-size effects) within certain limits.  By testing
the structures of the subsamples for several grid resolutions and values
of $\tau$, we chose $\tau=0.688$ for the main simulations of the
permeabilities. This is not necessarily the optimal choice and it
differs from the results for pipe flow, but a more
accurate calibration of $\tau$ would require further simulations or
other independent results. 

The permeability $k^{(\infty)}$ for an infinite resolution of the
lattice is obtained from a linear extrapolation of the data displayed in
Fig. \ref{fig:sub-k-a}. Extrapolation of the \FD results yields
$k^{(\infty)}=224\mD$. Thus, we obtain an estimate for the relative
error of our permeabilities in the case of \EX,
$(k^{(1)}-k^{(\infty)})/k^{(1)} \approx 0.33$. 
For \LB the extrapolated permeability is $k^{(\infty)}=550\mD$.
Analogously, for $\tau=0.688$ we obtain $(k^{(1)}-k^{(\infty)})/k^{(1)}
\approx 0.15$. For \DM this error of \LB is about the same size as for
\EX, and approximately $0.3-0.4$ for \SA and \GF. 

The simulations of the subsamples show also that the magnitude of
fluctuations in permeability, within one and the same sample, can be 100
\% even when the sample is extremely homogeneous. An effect of similar
size may be induced by inaccurate boundary conditions. This is important
when comparing two micropermeameter experiments in which different
boundary conditions may have applied.

When analyzing a subsample of sample \SA, we encountered another
difference between the \FD and \LB simulation. 
Geometrical analysis of the considered subsample of \SA reveals that 
this subsample is not percolating in direction $\vec e_1$. We found 
nevertheless that its \LB permeability is  $k_{11}^{(1)}=50 \mD$ 
for $\tau=0.688$, and for the original resolution $a_1=7.5 \mum$. 
We attribute this result to diagonal leaks that are present in the 
\LB model used.

\subsection{Velocity distributions}
\begin{figure}
  \setlength{\unitlength}{\linewidth}
  \begin{picture}(1,.7)(0,0)
    \put(0,0){\epsfig{file=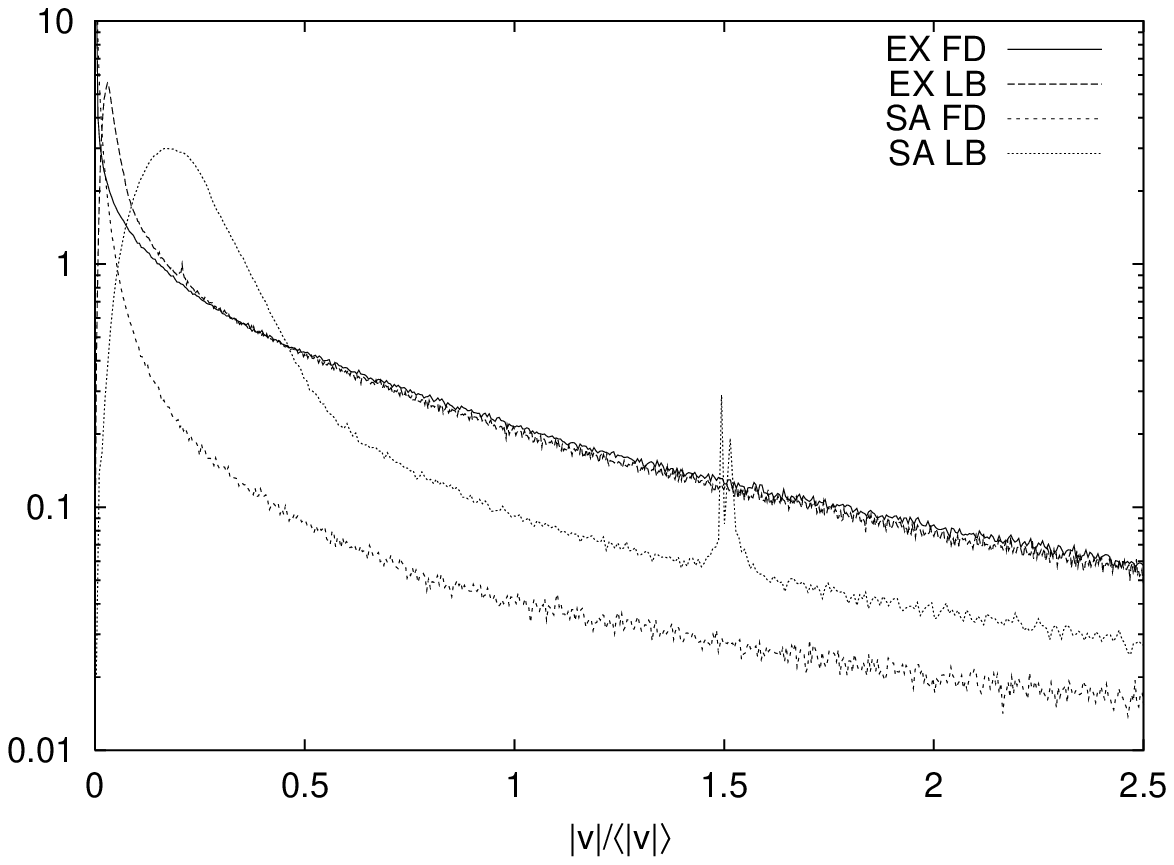, width=0.9\linewidth}}
    \put(0.35,0.35){\epsfig{file=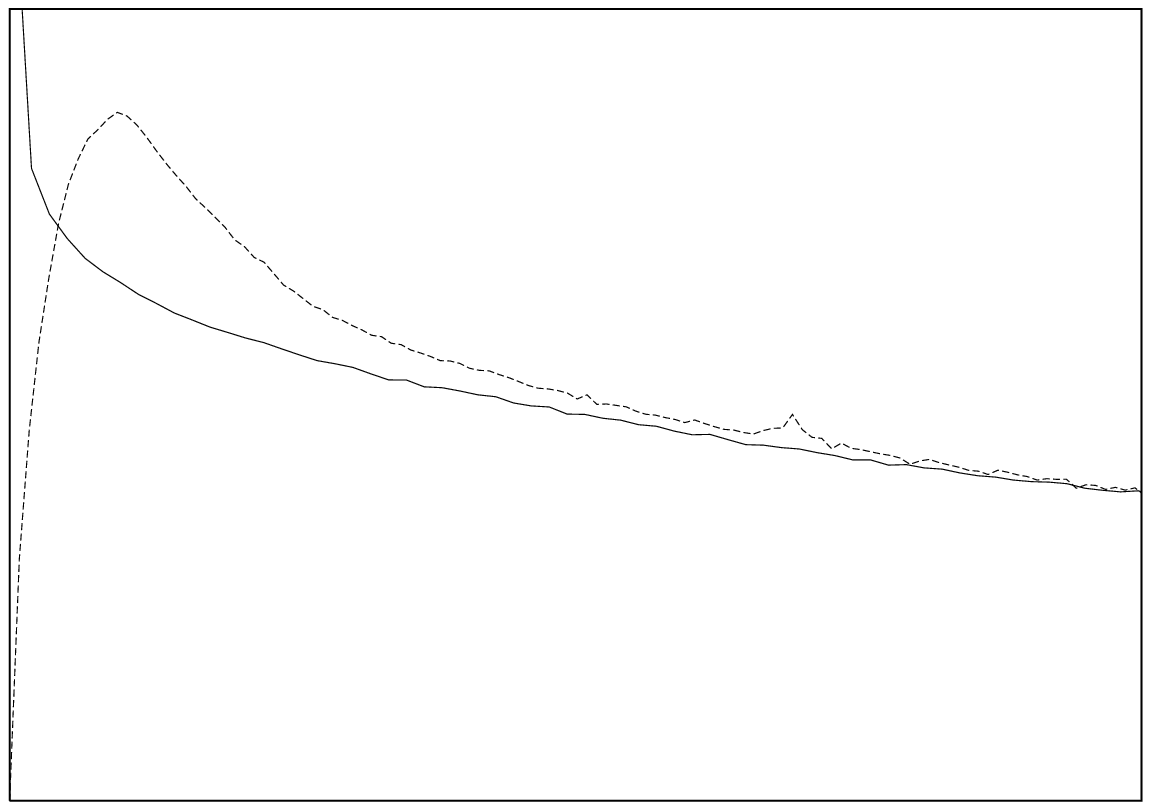, width=0.3\linewidth}}
  \end{picture}
  
  \caption{Velocity distribution function $P(|\vec v|)$ of the
    Fontainebleau sandstone (\EX) and the \SA model sampled over the
    pore space. The distributions are scaled with the mean velocity
    $\bra |\vec v(\vec x)| \ket_{\vec x \in \PPP}$. The inset shows a
    magnification of the distribution of \EX for small velocities. }
  \label{fig:sand-hist}
\end{figure}

Next we consider the velocity fields in more detail, and analyze the
histograms of velocity magnitudes. Figure \ref{fig:sand-hist} shows the
scaled distributions of the magnitude of the velocity $|\vec v(\vec x)|$
with $\vec x \in \PPP$ for samples \EX and \SA. The distributions were
sampled using the solutions of the flow fields for an applied pressure
gradient in direction $\vec e_1$. The higher permeability of the
original sandstone \EX is reflected by a higher probability density of
regions with average flow velocity. The distribution of sample \SA on
the other hand exhibits a higher peak at $|\vec v|=0$, and it extends to
higher velocities. The former observation indicates large stagnant areas
where no transport is taking place. The increased probability at high
velocities may be related to a large number of narrow pore throats
through which the fluid has to move. 
The velocity results for sample \GF resemble closely to those of
sample \SA while those for sample \DM resemble those for \EX.

Unlike the results for the permeabilities, which are close to
each other,
the velocity distributions reveal more significant differences
between \FD and \LB calculations (see Fig. \ref{fig:sand-hist}). 
The distributions obtained from the \LB
solution exhibit a maximum at small velocities which is not present in
the distributions obtained from the \FD solution. 
The differences could perhaps be attributed to the slip velocities 
at the boundary due, {\em e.g.}, to the bounce-back
boundary condition and the diagonal leak flows in the \LB
algorithm, which could lead to a systematic deficit of zero velocities
near the boundary.

For \SA the \LB simulations show an additional spurious 
(double) peak at $|\vec v|/\bra |\vec v| \ket \approx 1.5$. 
Similar, although smaller peaks were also found
for \GF and \EX. 
The locations of these peaks were found to be $\Delta P / 2 \rho
\bra |\vec v|\ket$, where $ \Delta P$ is the body force. 
We attribute these peaks to the staggered momenta found in small closed
pores, the number of which is very high for \SA and \GF. 
We have checked that these peaks disappear when time-averaged
velocities are used, and a corresponding increase appears at
zero velocity, which is the expected velocity for small closed 
pores.

\section{conclusion}

We have performed numerical micropermeametry on three dimensional porous
microstructures with a linear size of approximately 3 mm and a 
resolution of 7.5 $\mu$m. 
One of the samples has been a microtomographic image of Fontainebleau
sandstone. Two of the samples were stochastic reconstructions with
the same porosity, specific surface area, and two-point correlation
function as the Fontainebleau sample. 
The fourth sample was a physical model, which mimics the
processes of sedimentation, compaction and diagenesis of 
Fontainebleau sandstone.
The permeabilities of these samples were determined by numerically
solving at low Reynolds numbers the appropriate Stokes equations in the
pore spaces of the samples, using standard finite differences methods
and the lattice-Boltzmann method.
Our work shows that both methods, the \LB method as well as 
standard \FD methods, are applicable to the solution of the
steady-state Stokes equation within the microstructure of a
three-dimensional porous medium. 
We investigated systems with sizes of up
to $400^3$ lattice points. The solution of even larger systems seems
possible. Hence, numerical micropermeametry is becoming a feasible
technique for studying permeability fluctuations.

An accurate, quantitative comparison of the two numerical methods is
difficult due to the different approaches underlying these methods.
The memory requirements of the algorithms used in this study differ by
roughly a factor of 2.5. The \FD algorithm requires to store $8$ real
numbers per lattice node, while the \LB algorithm needs $19$ real
numbers per node.  
However, in this \LB model 15 real numbers could
also be used \cite{Qia92}.  
Considering the time consumption, both methods are quite similar.
Our comparison has shown that there are some features in the 
standard application of 
the \LB method, which need special attention.
These include the $\tau$-dependence of the no-flow boundary, the
compressibility of the fluid, staggered invariants, and diagonal leak
flows. One should notice that these occasionally inconvenient features
can usually be eliminated if so needed.  The compressibility of the
fluid can be eliminated for stationary flows \cite{Lin96}, and the
effects of staggered invariants can be eliminated with proper
averaging. The diagonal leak flow, which becomes noticeable for rough
surfaces and near the percolation threshold and limits there the
accuracy of the method, can also be eliminated by fairly
straightforward means. The $\tau$-dependence of the no-flow boundaries
can as well be eliminated, entirely by introducing a modified \LB
model \cite{Gin94}, or almost entirely by using second-order
boundaries as discussed above.  We could of course have implemented
here all these corrections in the \LB code, and achieved thereby a
more favourable comparison with the \FD code and experiment, but we
wanted to show the points of concern in a ``standard'' implementation of
the \LB method.

Our results provide a quantitative comparison of various models for
porous rocks.  We show for the first time that stochastic
reconstruction models for Fontainebleau are less accurate than
originally believed \cite{adl92,YT98a}.  In addition, our results for
the permeabilities of the Fontainebleau sandstone and its models
confirmed previous predictions \cite{hil99c} of a purely geometrical
investigation of the same samples based on local porosity theory. The
numerical value of $k$ for the \EX sample was found to be in good
agreement with the experimental value. Nevertheless, one has to keep
in mind that ``numerically exact'' results for the fluid permeability
must be handled with care. We have shown here that they may depend on
the numerical method, the boundary conditions, the size of the sample,
and the resolution of the microstructure.  Errors of as much as 100 \%
cannot typically been ruled out.  In summary, numerically exact
determination of permeability is difficult to achieve. On the other
hand, the same is true for the precision measurements in
an experiment.

\section*{Acknowledgments}
U.A., A.K. and J.T. thank the Center for Scientific Computing in Finland
for providing computational resources, and the Academy of Finland for
financial support under the MaDaMe Programme and the Finnish Center of
Excellence Programme 2000-2005 (Project No. 44875).
C.M. and R.H. thank the HLRS at Universit\"at Stuttgart for
providing computing time and the Deutsche Forschungsgemeinschaft
for financial support.


\clearpage
\begin{table}
  \begin{center}
    \setlength{\tabcolsep}{8pt}
    \begin{tabular}{l|>{$}r<{$}>{$}r<{$}>{$}r<{$}>{$}r<{$}}
       & \hbox{\EX} & \hbox{\DM} & \hbox{\GF} & \hbox{\SA} \\
      \hline
      $\bar\phi$   & 0.1355 & 0.1356 & 0.1421 & 0.1354 \\                       
      $S_V(\MMM)$\hfill  $[\mm^{-1}]$   & 9.99 & 10.30 & 14.53 & 11.04 \\
      $K_V(\MMM)$\hfill  $[\mm^{-2}]$ & -151 &-194   &-449   &-222   \\
      $T_V(\MMM)$\hfill  $[\mm^{-3}]$ & -2159 & -2766 & 4334 & 14484 \\
      $f_p$ [\%] & 99.35 & 99.23 & 79.16 & 62.73  \\ 
      $p_3(60a)$ & 0.9561 & 0.9647 & 0.3255 & 0.1695 \\
     \end{tabular}
  \end{center}
  \caption{Geometrical characteristics of the four samples. $S_V$,
    $K_V$ and $T_V$ are specific surface, specific integral of mean
    curvature and specific integral of total curvature of the matrix
    phase, respectively, $f_p$ is the fraction of percolating
    pore lattice points, and $p_3(60a)$ is the probability of finding a cubic subblock of
    size $L=60a$ of the sample, which is percolating in all
    three directions.}
  \label{tab:geo}
\end{table}

\begin{table}
  \begin{center}
    \setlength{\tabcolsep}{8pt}

\begin{tabular}{l|>{$}r<{$}>{$}r<{$}>{$}r<{$}|>{$}r<{$}>{$}r<{$}>{$}r<{$}}
      & & \hbox{\FD} & & & \hbox{\LB} & \\
      & k_{i1} & k_{i2}& k_{i3} & k_{i1} & k_{i2}& k_{i3} \\
      \hline
      EX &
       692  & 47  & -15 & 621 & 40  & -15 \\
      &15   & 911 &  50 & 14  & 808 & 47  \\
      &-103 & 21  & 789 & -85 & 15  & 687 \\
      \hline
      DM &
       923 &  40 &  16 & 766 & 31  & 10  \\
      & 27 & 581 &  25 & 19  & 482 & 22  \\
      & 21 &  35 & 623 & 14  & 32  & 528 \\
      \hline
      GF &
        34 &  1 &  4 & 43 &  3 &  6 \\
      &  0 & 35 &  2 & 1  & 50 &  6 \\
      &  8 &  1 & 36 & 7  & 4  & 57 \\
      \hline
      SA &
       35 &  0 &  5 & 56  & -2 & 7  \\
      &-7 & 22 & -1 & -5  & 46 & 13 \\
      & 3 & -7 & 20 & 8   & 1  & 50 \\
     \end{tabular}
  \end{center}    

  \caption{Permeability tensors of the Fontainebleau sandstone and its
    models. The values are given in ${\rm mD}$.
   }
  \label{tab:k}
\end{table}

\def\FDIR{./}

\begin{thebibliography}{10}

\bibitem{sah95}
M. Sahimi, {\em Flow and Transport in Porous Media and Fractured Rock} (VCH
  Verlagsgesellschaft mbH, Weinheim, 1995).

\bibitem{hil95d}
R. Hilfer, Advances in Chemical Physics {\bf XCII},  299  (1996).

\bibitem{KBDHLPS99}
P. King {\it et~al.}, Physica A {\bf 274},  60  (1999).

\bibitem{lak89}
L. Lake, {\em Enhanced Oil Recovery} (Prentice Hall, Englewood Cliffs, 1989).

\bibitem{Kop98}
A. Koponen, Ph.D. thesis, University of Jyv\"askyl\"a, 1998.

\bibitem{hel97}
R. Helmig, {\em Multiphase Flow and Transport Processes in the Subsurface}
  (Springer, Berlin, 1997).

\bibitem{hil99c}
B. Biswal {\it et~al.}, Physica A {\bf 273},  452  (1999).

\bibitem{hil94c}
R. Hilfer and P. {\O}ren, Transport in Porous Media {\bf 22},  53  (1996).

\bibitem{BZ85}
T. Bourbie and B. Zinszner, J.Geophys.Res. {\bf 90},  11524  (1995).

\bibitem{BCZ87}
T. Bourbie, O. Coussy, and B. Zinszner, {\em Acoustics of Porous Media}
  (Editions Technip, Paris, 1987).

\bibitem{dul92}
F. Dullien, {\em Porous Media - Fluid Transport and Pore Structure} (Academic
  Press, San Diego, 1992).

\bibitem{hil91d}
R. Hilfer, Phys. Rev. B {\bf 44},  60  (1991).

\bibitem{hil96}
R. Hilfer, Advances in Chemical Physics {\bf XCII},  299   (1996).

\bibitem{B:pat80}
S. Patankar, {\em Numerical Heat Transfer and Fluid Flow} (Hemisphere
  publishing corporation, New York, 1980).

\bibitem{hil00f}
A. Tscheschel {\it et~al.}, Physica A {\bf 284},  46  (2000).

\bibitem{hil00}
R. Hilfer,  in {\em Statistical physics and spatial statistics}, edited by K.
  Mecke and D. Stoyan (Springer, Berlin, 2000), Vol.~554, p.\ 203.

\bibitem{B:hir88}
C. Hirsch, {\em Numerical Calculation of Internal and External Flows} (Wiley \&
  Sons, New York, 1988), Vol.~1 \& 2.

\bibitem{HW65}
F. Harlow and J. Welsh, Physics of Fluids {\bf 8},  2182  (1965).

\bibitem{Qia92}
Y. Qian, D. d'Humi\'eres, and P. Lallemand, Europhys. Lett. {\bf 17},  479
  (1992).

\bibitem{Ben92}
R. Benzi, S. Succi, and M. Vergassola, Phys. Rep. {\bf 222},  145  (1992).

\bibitem{Rot97}
D. Rothman and S. Zaleski, {\em Lattice-Gas Cellular Automata: Simple Models of
  Complex Hydrodynamics}, {\em Collection Al{\'e}a} (Cambridge University
  Press, Cambridge, 1997).

\bibitem{CD98}
B. Chopard and M. Droz, {\em Cellular automata modelling of physical systems}
  (Cambridge University Press, Cambridge, 1998).

\bibitem{Fer95}
B. Ferr\'eol and D.~H. Rothman, Transport in Porous Media {\bf 20},  3  (1995).

\bibitem{Mar96}
N. Martys and H. Chen, Phys. Rev. E {\bf 53},  743  (1996).

\bibitem{Gal97}
M. Gallivan, D. Noble, J. Georgiadis, and R. Buckius, Int. J. Numer. Meth.
  Fluids {\bf 25},  249  (1997).

\bibitem{He97}
X. He, Q. Zou, L. Luo, and M. Dembo, J. Stat. Phys. {\bf 87},  115  (1997).

\bibitem{Kan99}
D. Kandhai {\it et~al.}, J. Comp. Phys. {\bf 150},  482  (1999).

\bibitem{Sko93}
P. Skordos, Phys. Rev. E {\bf 48},  4823  (1993).

\bibitem{Nob95}
D. Noble, S. Chen, J. Georgiadis, and R. Buckius, Phys. Fluids {\bf 7},  203
  (1995).

\bibitem{Fil98}
O. Filippova and D. H\"anel, Int. J. of Modern Phys. C {\bf 9},  1271  (1998).

\bibitem{Che98}
H. Chen, C. Teixeira, and K. Molvig, Int. J. of Modern Phys. C {\bf 9},  1281
  (1998).

\bibitem{Mai96}
R. Maier, R. Bernard, and D. Grunau, Phys. Fluids {\bf 8},  1788  (1996).

\bibitem{Che96}
S. Chen, D. Martinez, and R. Mei, Phys. Fluids {\bf 8},  2527  (1996).

\bibitem{Gin96}
I. Ginzbourg and D. d'Humier\'es, J. Stat. Phys. {\bf 84},  927  (1996).

\bibitem{Bui2000}
J. Buick and C. Greated, Phys. Rev. E {\bf 61},  5307  (2000).

\bibitem{Shan93}
X. Shan and H. Chen, Phys. Rev. E {\bf 47}, 1815 (1993)

\bibitem{Shan95}
X. Shan and H. Chen, J. Stat. Phys. {\bf 81}, 379 (1995)

\bibitem{Sch57}
A. Scheidegger, {\em The physics of flow in porous media} (The Macmillian
  company, New York, 1957).

\bibitem{Ver99}
R. Verberg and A.~J.~C. Ladd, Phys. Rev. E {\bf 60},  3366  (1999).

\bibitem{B:wie74}
K. Wieghardt, {\em Theoretische Str{\"o}mungslehre} (Teubner, Stuttgart, 1974).

\bibitem{LH89}
R. Larson and J. Higdon, Phys. Fluids A {\bf 1},  38   (1989).

\bibitem{lad88}
A. Ladd, J. Chem. Phys. {\bf 88},  5051   (1988).

\bibitem{man01}
C. Manwart, Ph.D. thesis, Universit{\"a}t Stuttgart, 2001.

\bibitem{Lin96}
Z. Lin, H. Fang, and R. Tao, Phys. Rev. E {\bf 54},  6323  (1996).

\bibitem{Gin94}
I. Ginzburg and P.~M. Adler, J. Phys. II (France) {\bf 4},  191  (1994).

\bibitem{adl92}
P. Adler, {\em Porous Media} (Butterworth-Heinemann, Boston, 1992).

\bibitem{YT98a}
C.~L.~Y. Yeong and S. Torquato, Physical Review E {\bf 57},  495  (1998).

\end{thebibliography}
\end{document}